\documentclass[
reprint,
superscriptaddress,
amsmath,
amssymb,
aps,
prl,
nobibnotes
]{revtex4-2}

\usepackage{array}
\usepackage{changes}
\usepackage{tabularx}
\usepackage{makecell}
\usepackage{graphicx}
\usepackage{dcolumn}
\usepackage{bm}
\usepackage{booktabs}
\usepackage{float}
\usepackage{placeins}
\usepackage{soul}
\usepackage{xcolor}
\usepackage[
    colorlinks=true,
    citecolor=blue,
    linkcolor=blue,
    urlcolor=blue
]{hyperref}

\begin{document}

\title{Coexistence of Altermagnetic and Compensated Ferrimagnetic Motifs in a Noncollinear Multiferroic}

\author{Vikrant Chaudhary}
\affiliation{Institute of Materials Science, Technical University of Darmstadt, Darmstadt 64287, Germany}
\affiliation{Physics Department and CSMB, Humboldt-Universität zu Berlin, 12489 Berlin, Germany}

\author{Fu Li}
\affiliation{Institute of Materials Science, Technical University of Darmstadt, Darmstadt 64287, Germany}

\author{Hui Liu}
\affiliation{Department of Physics, Stockholm University, AlbaNova University Center, 106 91 Stockholm, Sweden}

\author{Mengli Hu}
\affiliation{Leibniz Institute for Solid State and Materials Research Dresden, 01069 Dresden, Germany}

\author{Rajyavardhan Ray}
\affiliation{Department of Physics, Birla Institute of Technology Mesra, Ranchi, Jharkhand, India - 835215}

\author{Chen Shen}
\affiliation{Suzhou Laboratory, Suzhou 215021, China}

\author{Jeroen van den Brink}
\affiliation{Leibniz Institute for Solid State and Materials Research Dresden, 01069 Dresden, Germany}
\affiliation{W{\"u}rzburg-Dresden Cluster of Excellence ctd.qmat, 01069 Dresden, Germany}
\affiliation{Institute of Theoretical Physics, Technische Universit{\"a}t Dresden, 01062 Dresden, Germany}

\author{Hongbin Zhang}
\affiliation{Institute of Materials Science, Technical University of Darmstadt,
Darmstadt 64287, Germany}

\author{Harish K. Singh}
\email{harish.singh@plus.ac.at}
\affiliation{Department of Chemistry and Physics of Materials,
University of Salzburg, Jakob-Haringer-Strasse 2a,
5020 Salzburg, Austria}


\begin{abstract}
Collinear compensated magnets with nonrelativistic spin splitting comprise altermagnets (ALMs), whose rotation- or mirror-related opposite-spin sublattices produce momentum-selective splitting that vanishes at $\Gamma$, and compensated ferrimagnets (CFIMs), whose symmetry-inequivalent sublattices allow splitting throughout the Brillouin zone, including at $\Gamma$. Here, we show that the coplanar noncollinear magnetic state of multiferroic boracite Ni$_3$B$_7$O$_{13}$Cl hosts both motifs on distinct Ni sublattices in real space. The electronic structure and spin texture exhibit coexisting $d$-wave-like and $s$-wave-like spin-polarization components associated with the ALM and CFIM motifs, respectively. Opposite ferroelectric states exhibit reversed band-resolved spin polarizations, indicating magnetoelectric coupling, and sign reversals of linear shift and circular injection currents.
\end{abstract}

\maketitle
{\it Introduction--}
Altermagnetism represents a symmetry-distinct class of collinear compensated magnetic order that supports nonrelativistic spin splitting (NRSS) without net magnetization~\cite{hayami2019momentum,vsmejkal2020crystal,yuan2021prediction,mazin2021prediction,guo2023spin}. In conventional collinear antiferromagnets, a $180^\circ$ spin-space rotation combined with inversion or translation relates the opposite-spin sublattices and enforces spin degeneracy throughout the Brillouin zone, whereas in altermagnets (ALMs), the same spin rotation combined with a crystal rotation or mirror relates the sublattices, producing momentum-selective spin splitting of even-parity $d$-, $g$-, or $i$-wave form while retaining spin degeneracy at the $\Gamma$ point~\cite{vsmejkal2022beyond}. This characteristic band structure has been experimentally confirmed~\cite{krempasky2024altermagnetic,reimers2024direct} and can enable unconventional spintronic phenomena~\cite{vsmejkal2022emerging,gonzalez2021efficient,bai2024altermagnetism}.

Alongside altermagnets, compensated ferrimagnets (CFIMs) provide a complementary route to NRSS by combining zero net magnetization with magnetically inequivalent opposite-spin sublattices that are not related by any spin-interchanging symmetry, thereby placing CFIMs in a ferromagnetic-like symmetry class~\cite{yuan2024nonrelativistic,guo2025luttinger}. This symmetry allows $s$-wave NRSS throughout the Brillouin zone, including at the $\Gamma$ point, despite vanishing net magnetization, making CFIMs attractive for spintronic applications~\cite{zhou2021efficient,finley2020spintronics,semboshi2022new,lee2025spin,park2022anomalous,han2026strain}. CFIMs span two-dimensional materials~\cite{guo2025achieving,liu2025two,zhang2025fully,liu2025robust,guo2026electric,guo2026route,guo2026external}, bulk crystals~\cite{yuan2024nonrelativistic,guo2025luttinger,dong2025nonrelativistic,zhang2025prototypes}, and organic compounds~\cite{kawamura2024compensated}, with persistent nearly compensated ferrimagnetism observed above room temperature~\cite{aribot2026persistent,mavragani2026compensated}.

Although the symmetry classifications discussed above were developed mainly for collinear magnets with a common spin-quantization axis, recent studies have extended altermagnetism to noncollinear order~\cite{cheong2024altermagnetism,hu2025spin,hu2026non}. In magnets with several magnetic sublattices, local moments can align along different axes, forming a noncollinear magnetic structure. Can such a structure host both ALM and CFIM motifs in real space, and how would their coexistence govern the spin-splitting wave symmetry in momentum space? Here, we show that the noncentrosymmetric boracite Ni$_3$B$_7$O$_{13}$Cl~\cite{ye1997neutron}, a member of the boracite family recognized among the first multiferroics, hosts both motifs within a noncollinear magnetic state.

\begin{figure}[!htp]
        \centering
        \includegraphics[width=1.0\linewidth]{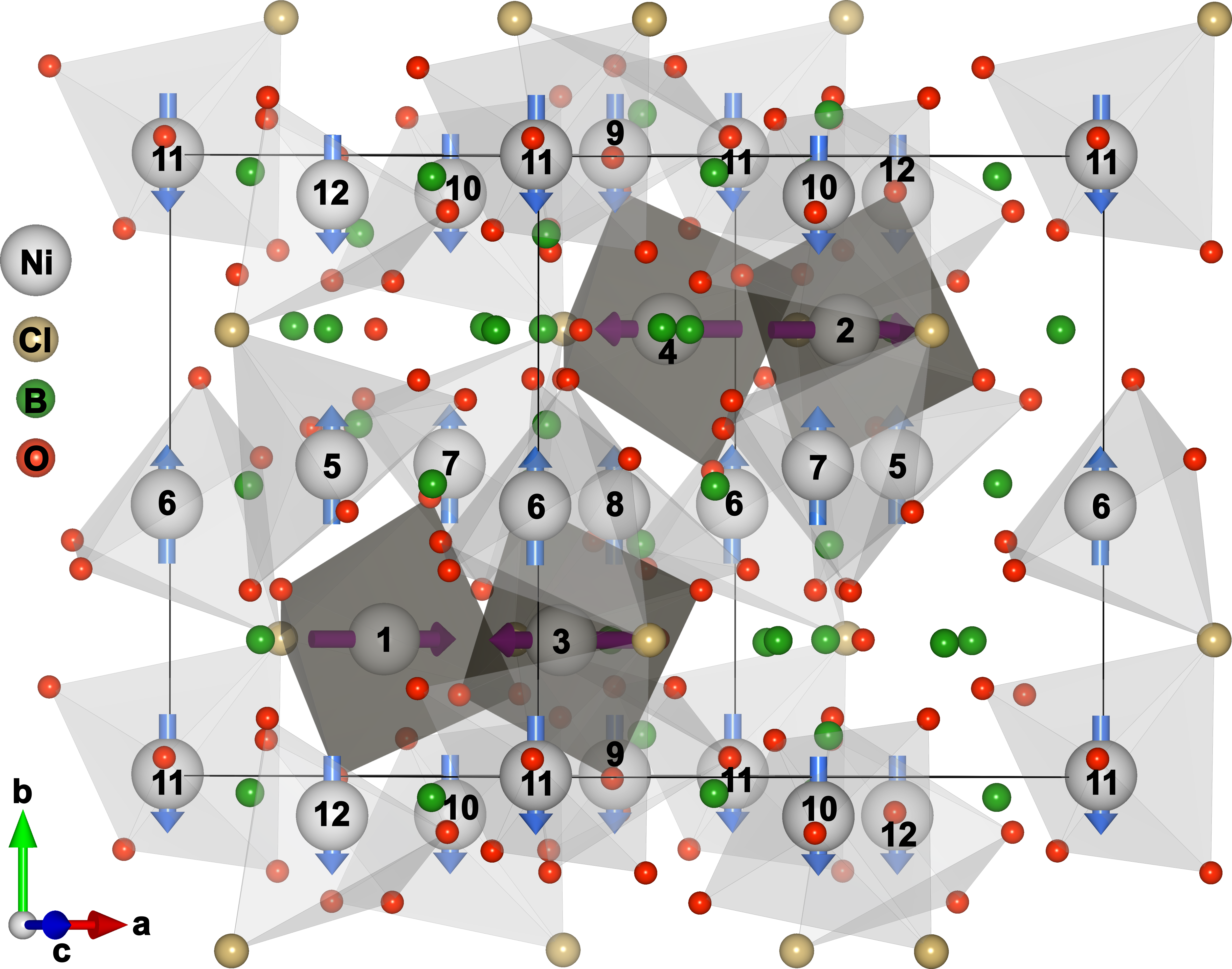}
        \caption{Crystal and magnetic structure of Ni$_3$B$_7$O$_{13}$Cl in the orthorhombic space group $Pca2_1$ in the $-P_z$ polarization state. Dark-gray polyhedra highlight the Ni-I sublattice (1--4), which forms the ALM motif, while light-gray polyhedra highlight the Ni-II (5--8) and Ni-III (9--12) sublattices, which form the CFIM motif. Arrows indicate the magnetic-moment directions, and numbers identify the Ni sites used in Table~\ref{tab:ssg_msg_ops}. The Cartesian axes are chosen as $x\parallel a$, $y\parallel b$, and $z\parallel c$.}
        \label{fig:crystal-magnetic-structure}
\end{figure}

In noncentrosymmetric compensated magnets, combining ferroelectricity with spin-polarized electronic states, as recently reported in both ALMs~\cite{gu2025ferroelectric} and CFIMs~\cite{dong2025nonrelativistic}, is particularly attractive because magnetoelectric coupling can enable electrical control of the magnetic order and associated spin-polarized bands while retaining negligible stray fields~\cite{eerenstein2006multiferroic}. Moreover, the second-order bulk photovoltaic effect (BPVE) requires broken inversion symmetry and generates a direct current under homogeneous optical excitation without external bias or a $p$--$n$ junction through two intrinsic contributions, shift current (SC), arising from a real-space displacement of the charge center during interband excitation, and injection current (IC), originating from an asymmetric population of photocarriers with unequal velocities~\cite{zhao2026recent,kim2017shift,panday2019injection}. Recent studies of nonlinear photocurrents in ALMs~\cite{ezawa2025bulk,dong2025crystal,jiang2025nonlinear,wang2026strain,li2026medium,li2026MagneticPhotocurrents} motivate exploring these responses in multiferroics with coexisting ALM and CFIM motifs.

In this work, we investigate the noncollinear multiferroic boracite Ni$_3$B$_7$O$_{13}$Cl. Using magnetic-space-group (MSG) and spin-space-group (SSG) analyses, we map the Ni sites to characterize the coexistence of ALM and CFIM motifs and examine how these real-space motifs govern spin splitting and the momentum-resolved spin polarization of the electronic bands. We extend this analysis using a minimal symmetry-allowed Hamiltonian to interpret the calculated band splitting in terms of coexisting $s$- and $d$-wave contributions and explain the finite splitting at $\Gamma$. We compare the two ferroelectric states polarized along $-z$ and $+z$, denoted $-P_z$ and $+P_z$, respectively, to probe magnetoelectric coupling between the polar state and the electronic spin texture and evaluate their linear SC and circular IC responses.

{\it Methods--}
Spin-polarized noncollinear density-functional theory (DFT) calculations, with and without spin--orbit coupling (SOC), were performed using the Vienna \textit{ab initio} simulation package (VASP)~\cite{kresse1993abinitio,kresse1994abinitio,kresse1996efficient,kresse1996efficiency}
within the projector-augmented-wave (PAW) formalism~\cite{blochl1994,kresse1999}. Exchange--correlation effects were treated using the Perdew--Burke--Ernzerhof (PBE) generalized-gradient approximation~\cite{perdew1996generalized}. The complete computational details, including the point-charge model for polarization, Wannier function calculations, and the formalisms for the photocurrents, are provided in Section~S1 of the Supporting Information (SI).

\begin{table}[t]
    \caption{Site mappings (a) MSG and (b) SSG operations for
Ni$_3$B$_7$O$_{13}$Cl. The operations act on representative sites Ni1, Ni5, and Ni9 of sublattices Ni-I, Ni-II, and Ni-III, respectively, with site labels defined in Fig.~\ref{fig:crystal-magnetic-structure}. Ni-I forms the ALM motif, while Ni-II and Ni-III jointly form the CFIM motif.}
    \label{tab:ssg_msg_ops}
    \centering
    \renewcommand{\arraystretch}{1.2}
    \setlength{\tabcolsep}{3pt}

    {(a) MSG: $\mathrm{Pc^{\prime}a2_{1}^{\prime}}$
    (No.~29.101).}\par\smallskip

    \begin{tabular}{|c|c|c|c|}
        \hline
        MSG & Ni-I & Ni-II & Ni-III \\
        Symmetry & Spin & Spin & Spin \\
        operations & sublattice & sublattice & sublattice \\
        & (Ni1) & (Ni5) & (Ni9) \\
        \hline
        $\{1 \mid 0\}$
        & $1\rightarrow1$ & $5\rightarrow5$ & $9\rightarrow9$ \\

        $\{C_{2z}^{\prime} \mid 0\,0\,\tfrac12\}$
        & $1\rightarrow2$ & $5\rightarrow6$ & $9\rightarrow10$ \\

        $\{m_x^{\prime} \mid \tfrac12\,0\,\tfrac12\}$
        & $1\rightarrow3$ & $5\rightarrow7$ & $9\rightarrow11$ \\
        
        $\{m_y \mid \tfrac12\,0\,0\}$
        & $1\rightarrow4$ & $5\rightarrow8$ & $9\rightarrow12$ \\
        \hline
    \end{tabular}

    \vspace{0.8em}

    {(b) SSG:
    $\mathrm{P}^{2_{010}}\mathrm{c}^{2_{010}}
    \mathrm{a}^{1}2_{1}^{m_{001}}1$.}\par\smallskip

    \begin{tabular}{|c|c|c|c|}
        \hline
        SSG & Ni-I & Ni-II & Ni-III \\
        Symmetry & Spin & Spin & Spin \\
        operations & sublattice & sublattice & sublattice \\
        & (Ni1) & (Ni5) & (Ni9) \\
        \hline
        $[\,1\!\parallel\!1 \mid 0\,0\,0\,]$
        & $1\rightarrow1$ & $5\rightarrow5$ & $9\rightarrow9$ \\

        $[\,1\!\parallel\!C_{2z} \mid 0\,0\,\tfrac12\,]$
        & $1\rightarrow2$ & $5\rightarrow6$ & $9\rightarrow10$ \\

        $[\,C_{2y}\!\parallel\!m_x \mid \tfrac12\,0\,\tfrac12\,]$
        & $1\rightarrow3$ & $5\rightarrow7$ & $9\rightarrow11$ \\
        
        $[\,C_{2y}\!\parallel\!m_y \mid \tfrac12\,0\,0\,]$
        & $1\rightarrow4$ & $5\rightarrow8$ & $9\rightarrow12$ \\

        $\{\,m_{z}\!\parallel\!1 \mid 0\,0\,0\,\}$ & $1\rightarrow1$ & $5\rightarrow5$  & $9\rightarrow9$  \\

        $\{\,m_{z}\!\parallel\!C_{2z} \mid 0\,0\,\tfrac12\,\}$ & $1\rightarrow2$ & $5\rightarrow6$ & $9\rightarrow10$  \\

        $\{\,m_{x}\!\parallel\!m_{x} \mid \tfrac12\,0\,\tfrac12\,\}$ & $1\rightarrow3$ & $5\rightarrow7$ & $9\rightarrow11$  \\

        $\{\,m_{x}\!\parallel\!m_{y} \mid \tfrac12\,0\,0\,\}$ & $1\rightarrow4$ &$5\rightarrow8$  &$9\rightarrow12$  \\
        \hline
    \end{tabular}
\end{table}

{\it Crystal and magnetic structures--}
Ni$_3$B$_7$O$_{13}$Cl crystallizes in the polar orthorhombic space group $Pca2_1$ (No.~29) below 610~K~\cite{ye1997neutron}. Each Ni$^{2+}$ ion is coordinated by four oxygen atoms and one chlorine atom, forming a distorted square-pyramidal NiO$_4$Cl polyhedron, with three neighboring polyhedra sharing a Cl vertex. Below $T_N\approx9$~K, neutron diffraction reveals a coplanar noncollinear spin configuration, which we adopt in this study~\cite{ye1997neutron}. The refined structure belongs to the magnetic space group $Pc^\prime a2_1^\prime$ (29.101), with magnetic point group $m^\prime m2^\prime$ (7.3.22). The three inequivalent Ni sublattices occupy $4a$ Wyckoff sites, with Ni-I (1--4) moments collinear and antiparallel along $a$ and Ni-II (5--8) and Ni-III (9--12) moments collinear along $b$, parallel within each sublattice and antiparallel between them (Fig.~\ref{fig:crystal-magnetic-structure}).

Nevertheless, single-domain crystals exhibit a weak moment of $0.064~\mu_{\mathrm B}$/f.u.\ along $b$~\cite{haida1974}, while polycrystals show $0.023~\mu_{\mathrm B}$/f.u.~\cite{schnelle2015magnetic}. For the $-P_z$ and $+P_z$ states (\textcolor{blue}{Fig.~S1}), noncollinear calculations with (without) SOC yield net moments along $b$ of $0.069$ ($0.004$) and $0.074$ ($0.002$)~$\mu_{\mathrm B}$/f.u., respectively. The SOC values are comparable to experiment and indicate nearly compensated magnetism. A small residual magnetization does not preclude fast spin dynamics, as ferrimagnets can exhibit antiferromagnetic-like dynamics near angular-momentum compensation~\cite{stanciu2006ultrafast,kim2017fast,caretta2018fast}.

To identify the magnetic motifs, we examine how MSG and SSG operations map the representative sites Ni1, Ni5, and Ni9 of Ni-I, Ni-II, and Ni-III, respectively (Table~\ref{tab:ssg_msg_ops}). The MSG glide operations $\{m_x^\prime\thinspace |\thinspace \frac12\thinspace 0\thinspace \frac12\}$ and $\{m_y\thinspace |\thinspace \frac12\thinspace 0\thinspace 0\}$ map Ni1 onto Ni3 and Ni4, respectively, connecting the antiparallel $a$-directed moments within Ni-I. Here, $m_x$ and $m_y$ denote mirrors normal to $a$ and $b$, respectively, and the prime denotes time reversal. In the nonrelativistic description, an SSG operation $[U\parallel R\thinspace|\thinspace\boldsymbol{\tau}]$ separates the spin transformation $U$ from the real-space operation $\{R\thinspace|\thinspace\boldsymbol{\tau}\}$. The operations $[C_{2y}\parallel m_y|\frac12\thinspace 0\thinspace 0]$ and $[C_{2y}\parallel m_x|\frac12\thinspace 0\thinspace \frac12]$ reproduce these mappings, with the $180^\circ$ spin rotation $C_{2y}$ reversing the $a$-directed moments while preserving the $b$-directed moments (Table~\ref{tab:ssg_msg_ops}). These symmetry-related antiparallel moments form an ALM motif within Ni-I. In contrast, both analyses map the sites of Ni-II and Ni-III only within their respective sublattices, with no operation interchanging them (Table~\ref{tab:ssg_msg_ops}). These moments along $+b$ and $-b$ therefore form a CFIM motif, establishing the coexistence of both motifs within the noncollinear magnetic structure.

\begin{figure}[!htp]
\centering
\includegraphics[width=1.0\linewidth]{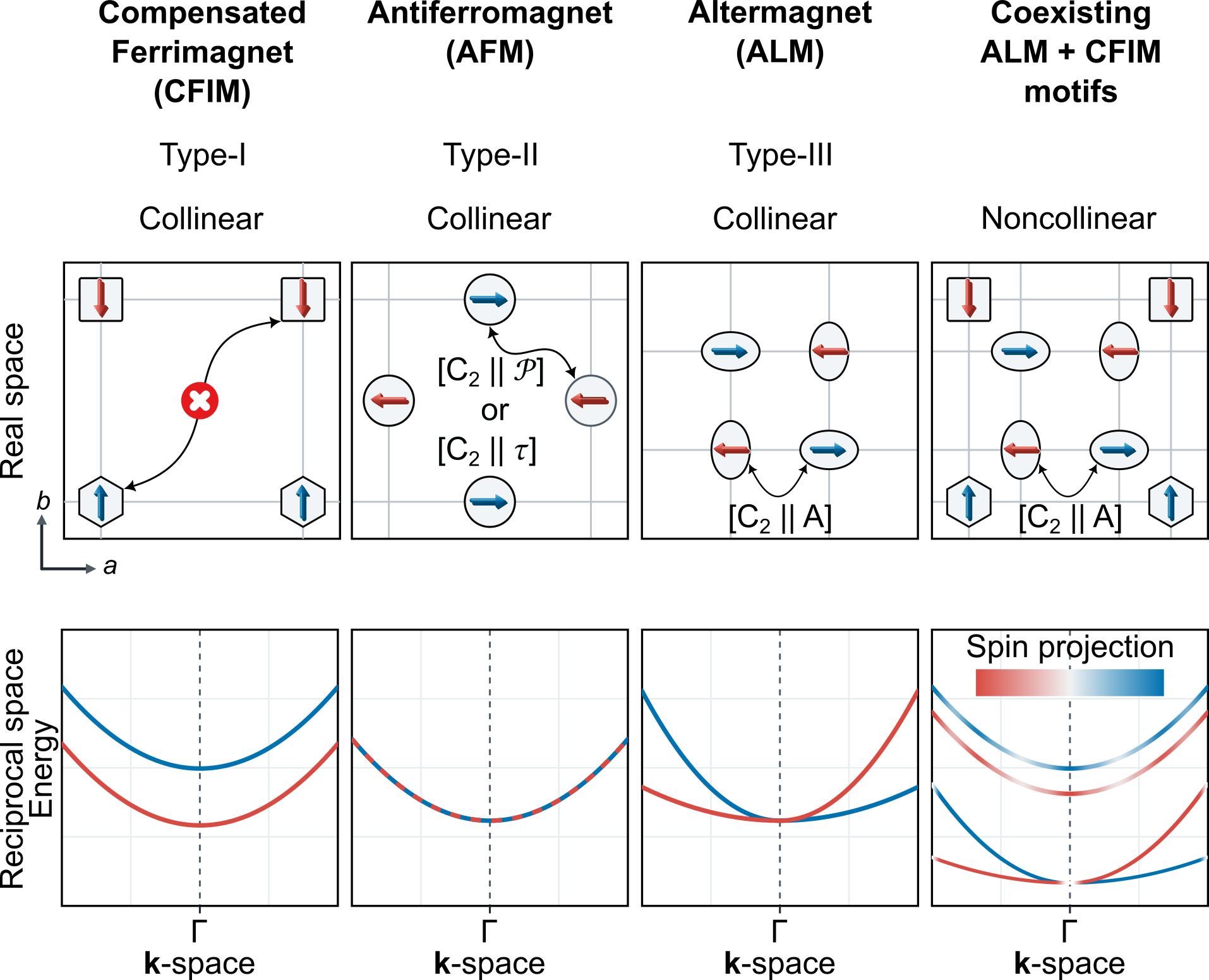}
\caption{Schematic real-space sublattice relations (top) and band structures (bottom) for collinear CFIM (Type I), AFM (Type II), and ALM (Type III) states, and a noncollinear state with coexisting ALM and CFIM motifs. Here, $C_2$ denotes a $180^\circ$ spin rotation, $\mathcal{P}$ inversion, $\tau$ translation, and $A$ a crystal rotation or mirror. Spin arrows are overlaid on the schematic sites for clarity and do not imply spin--lattice locking in the spin-group description. Band splittings are schematic and not to scale.}
\label{fig:classification}
\end{figure}

{\it Symmetry classification--}
As illustrated in Fig.~\ref{fig:classification}, the nonrelativistic collinear spin-Laue classification assumes a common spin-quantization axis and connects the real-space relations between opposite-spin sublattices to spin degeneracy or splitting in momentum space~\cite{vsmejkal2022beyond}. Type-I groups, $R_s^{\mathrm{I}}=[E\thinspace\Vert\thinspace G]$, include CFIMs, in which no spin-interchanging symmetry relates the opposite-spin sublattices, and allow spin splitting throughout the Brillouin zone, including at $\Gamma$. Type-II groups, $R_s^{\mathrm{II}}=[E\thinspace\Vert\thinspace G]+[C_2\thinspace\Vert\thinspace G]$, describe conventional antiferromagnets, in which $[C_2\thinspace\Vert\thinspace\mathcal{P}]$ or $[C_2\thinspace\Vert\thinspace\tau]$ relates opposite-spin sublattices through inversion or translation and enforces spin degeneracy throughout the Brillouin zone. Type-III groups, $R_s^{\mathrm{III}}=[E\thinspace\Vert\thinspace H]+[C_2\thinspace\Vert\thinspace G-H]$, describe ALMs, in which $[C_2\thinspace\Vert\thinspace A]$ relates opposite-spin sublattices through a crystal rotation or mirror, producing momentum-selective spin splitting while retaining degeneracy at $\Gamma$. Here, $E$ denotes the spin-space identity, $C_2$ a spin-reversing $180^\circ$ rotation, $G$ the crystallographic Laue group, and $H$ the halving subgroup of spin-preserving spatial operations.

In a noncollinear magnet, however, the effect of a spin rotation depends on the moment direction. For Ni$_3$B$_7$O$_{13}$Cl, spin-space $C_{2y}$ transforms $(\sigma_x,\sigma_y,\sigma_z)$ into $(-\sigma_x,\sigma_y,-\sigma_z)$. The operations $[C_{2y}\thinspace\Vert\thinspace m_x]$ and $[C_{2y}\thinspace\Vert\thinspace m_y]$ therefore connect opposite $x$-directed moments within Ni-I while preserving the $y$-directed moments within Ni-II and Ni-III (Table~\ref{tab:ssg_msg_ops}). Thus, the same spin rotation reverses the spin component associated with the ALM motif but preserves that associated with the CFIM motif. The noncollinear state therefore cannot be assigned directly to a collinear Type-I or Type-III class. 

The coexistence of the ALM and CFIM motifs cannot be maintained when all Ni moments are made collinear along either the $x$ or $y$ axis while retaining their relative sign pattern. In these configurations, the glide-related SSG symmetries are absent, and no SSG operation relates the opposite-spin Ni-I sites. The twofold screw along $z$ is paired with the spin-space identity and connects only same-spin sites. Therefore, noncollinearity is essential for realizing these coexisting ALM and CFIM motifs.

{\it Minimal effective Hamiltonian--}
The electronic consequences of these symmetry relations can be illustrated by a minimal symmetry-allowed two-state Hamiltonian in the SOC-free limit near the $\Gamma$ point, consistent with the similarity of the band structures obtained with and without SOC (\textcolor{blue}{Fig.~S2}). Retaining a constant $y$-spin term and the leading symmetry-allowed momentum-dependent $x$-spin term gives
\begin{equation}
H_{\mathrm{eff}}(\mathbf{k})
= \varepsilon_0(\mathbf{k})\mathbf{I}_2
+ \Delta_s\sigma_y
+ \lambda k_xk_y\sigma_x,
\label{eq:effective_spin_hamiltonian}
\end{equation}
\begin{figure}[!htp]
        \centering
        \includegraphics[width=1.0\linewidth]{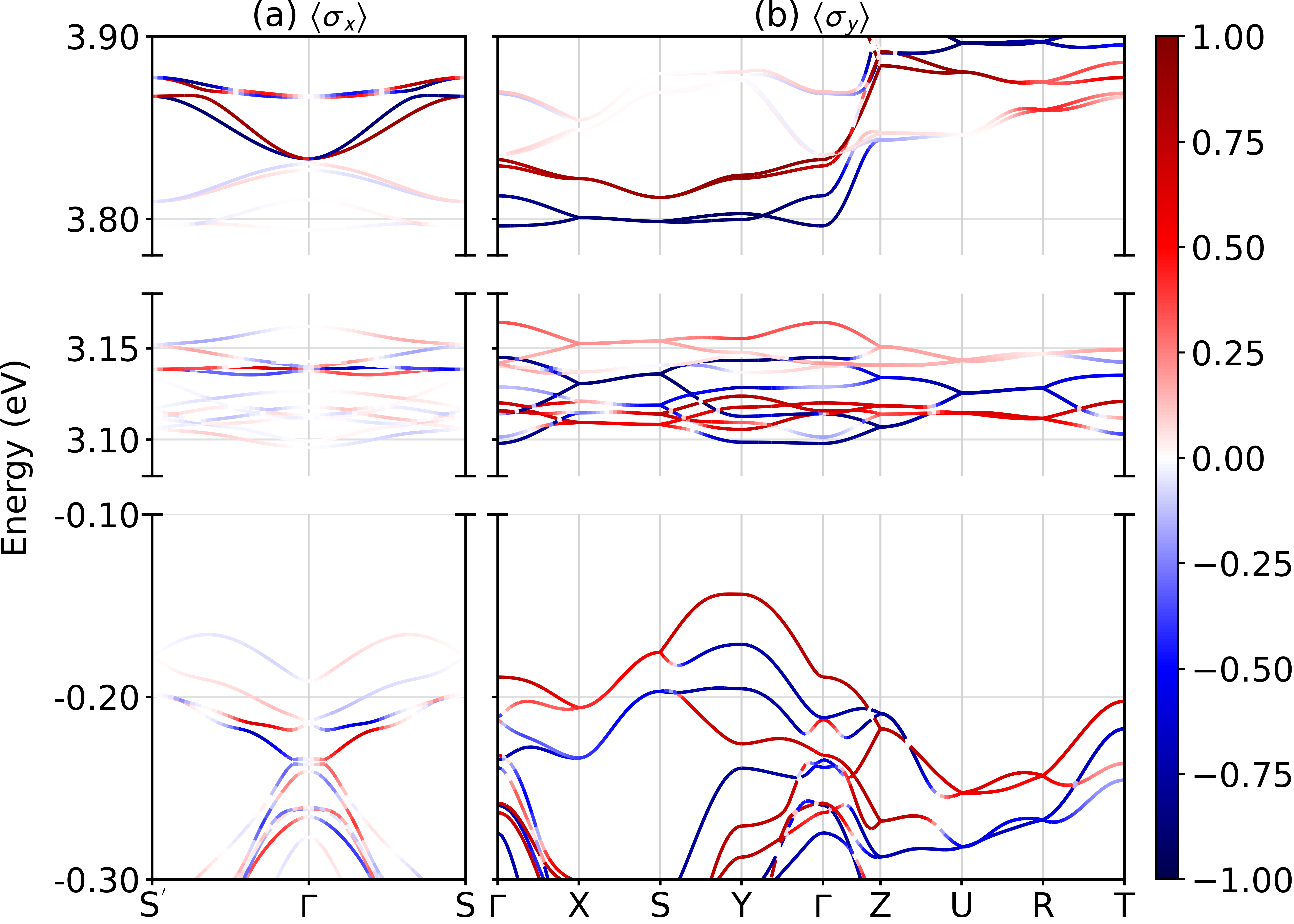}
        \caption{Noncollinear band structures of the $-P_z$ state calculated without SOC, projected onto (a) $\langle\sigma_x\rangle$ and (b) $\langle\sigma_y\rangle$. The color bar indicates the $x$-spin polarization $\langle\sigma_x\rangle$ and $y$-spin polarization $\langle\sigma_y\rangle$. The Brillouin zone and high-symmetry points are shown in \textcolor{blue}{Fig.~S3}, where $S'=(-\tfrac{1}{2},\tfrac{1}{2},0)$. Red and blue indicate positive and negative spin projections, respectively. The Fermi energy is set to zero. Two breaks in the energy axis omit band-free intervals for clarity.}
        \label{fig:band-reversal}
\end{figure}
where $\varepsilon_0(\mathbf{k})$ is the spin-independent dispersion, $\mathbf{I}_2$ is the $2\times2$ identity matrix, and $\sigma_x$ and $\sigma_y$ are Pauli matrices acting in the model spin space. The parameter $\Delta_s$ is the coefficient of the momentum-independent $s$-wave contribution, while $\lambda$ controls the strength of the momentum-dependent $d$-wave contribution. The term $\lambda k_xk_y\sigma_x$ is symmetry-allowed because, under either combined mirror operation, $[C_{2y}\thinspace\Vert\thinspace m_x]$ or $[C_{2y}\thinspace\Vert\thinspace m_y]$, both $k_xk_y$ and $\sigma_x$ reverse sign, leaving their product unchanged. Both retained terms also respect the coplanar spin-only antiunitary symmetry $[m_z^{\mathrm{s}}\thinspace\Vert\thinspace E]$, under which momentum reverses while the in-plane spin components remain unchanged. Symmetry-allowed quadratic contributions proportional to $k_i^2\sigma_y$ ($i=x,y,z$), which have the same momentum order as the retained $x$-spin term, and the leading cubic contribution proportional to $k_xk_yk_z\sigma_z$ are omitted for simplicity. At the $\Gamma$ point, where $\mathbf{k}=0$, the $d$-wave contribution vanishes, yielding
\begin{equation}
H_{\mathrm{eff}}(\Gamma)
= \varepsilon_0(\Gamma)\mathbf{I}_2
+ \Delta_s\sigma_y.
\label{eq:gamma_spin_splitting}
\end{equation}
Therefore, finite splitting remains at $\Gamma$ whenever $\Delta_s\neq0$. The allowed $s$-wave contribution illustrates how CFIM-like finite splitting at $\Gamma$ can coexist with the embedded ALM motif, while the full magnetic state retains its noncollinear spin-group symmetry. 

\begin{figure}[!htp]
        \centering
        \includegraphics[width=1.0\linewidth]{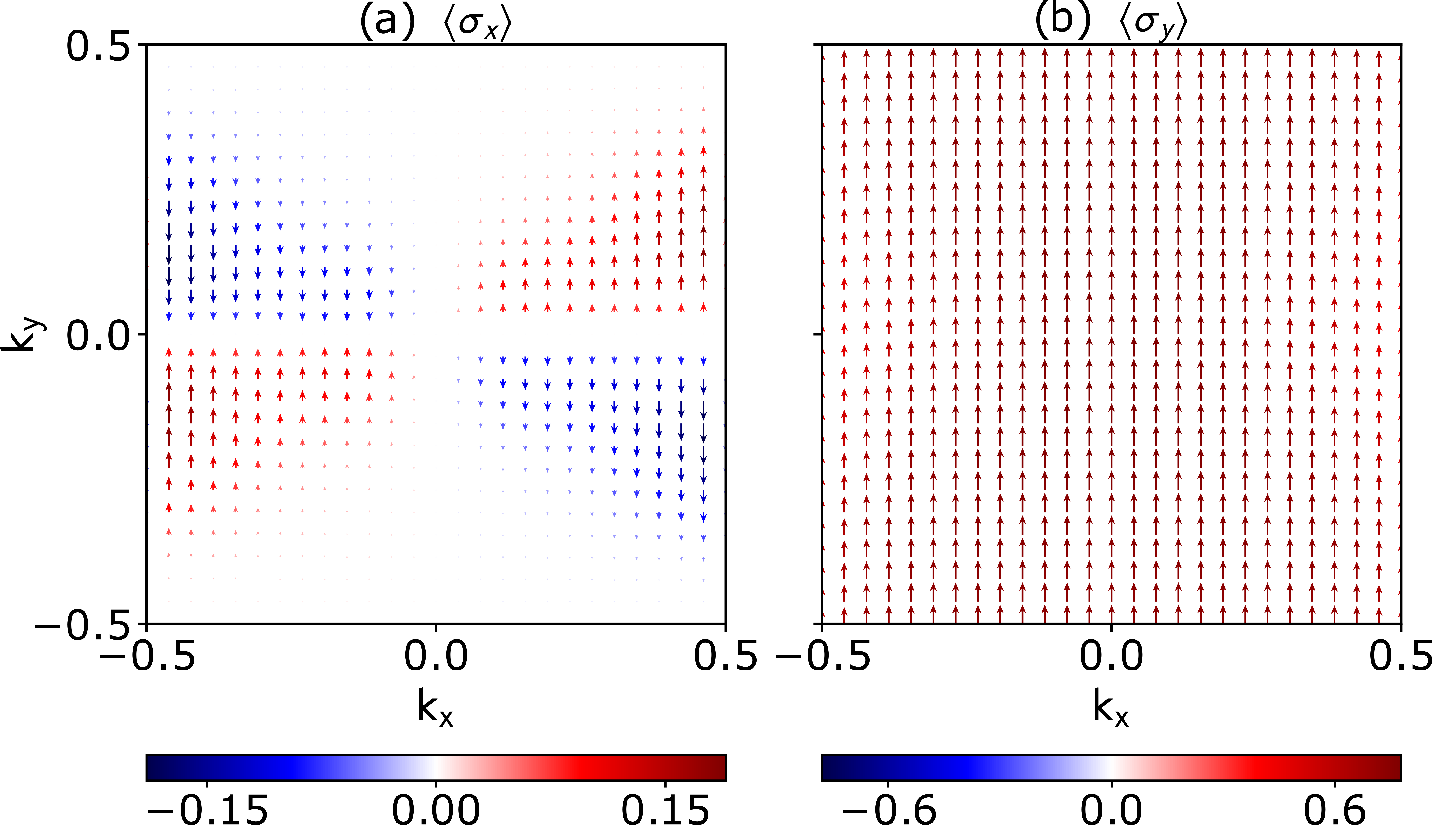}
        \caption{Spin texture of the valence-band maximum (VBM) in the $k_x$–$k_y$ plane for the $-P_z$ state without SOC, showing (a) $d$-wave-like $\langle\sigma_x\rangle$ and (b) $s$-wave-like $\langle\sigma_y\rangle$. Colors indicate projection values, with upward (downward) arrows denoting positive (negative) signs.}
        \label{fig:spin-texture}
\end{figure}

{\it Band structure--}
These real-space symmetry relations are reflected in the electronic structure. The noncollinear bands calculated without SOC exhibit spin splitting throughout the Brillouin zone, including at $\Gamma$. The $x$-spin polarization $\langle\sigma_x\rangle$ reveals $d$-wave-like behavior associated with the ALM motif (Fig.~\ref{fig:band-reversal}(a)), whereas the $y$-spin polarization $\langle\sigma_y\rangle$ is $s$-wave-like, consistent with CFIM character (Fig.~\ref{fig:band-reversal}(b)). These complementary signatures are also evident in the spin texture of the band forming the valence-band maximum (VBM), where $\langle\sigma_x\rangle$ alternates in sign between adjacent quadrants while $\langle\sigma_y\rangle$ preserves its sign (Fig.~\ref{fig:spin-texture}). At $\Gamma$, symmetry enforces $\langle\sigma_x\rangle=0$, while the finite band splitting is consistent with the $s$-wave contribution in the minimal Hamiltonian.

To further verify the presence of an ALM motif, we constructed a hypothetical reference structure by replacing the Ni atoms on the Ni-II and Ni-III sublattices with nonmagnetic Zn. The resulting bands exhibit the momentum-selective spin splitting characteristic of an ALM, remaining spin-degenerate at the $\Gamma$ point and along certain high-symmetry paths while splitting along others (\textcolor{blue}{Fig.~S4}). This comparison supports the identification of an embedded ALM motif within Ni-I.

The $y$-spin component $\langle\sigma_y\rangle$ is symmetry-allowed throughout the Brillouin zone both with and without SOC (\textcolor{blue}{Table~S1}). Among the examined paths, finite $\langle\sigma_x\rangle$  without SOC is restricted to the interiors of $S^\prime$--$\Gamma$--$S$. With SOC, finite $\langle\sigma_x\rangle$ also appears on selected bands in the interiors of $Z$--$T$, $U$--$R$, $X$--$S$, and $Y^\prime$--$\Gamma$--$Y$ (\textcolor{blue}{Fig.~S5}), consistent with the symmetry constraints. At wave vectors invariant under the unitary glide $\{m_y\thinspace|\thinspace\tfrac12\thinspace0\thinspace0\}$, nondegenerate Bloch states satisfy $\langle\sigma_x\rangle=\langle\sigma_z\rangle=0$, allowing only $y$-directed spin polarization. Away from these wave vectors, additional spin components are allowed depending on the remaining symmetries (\textcolor{blue}{Table~S1}).

To quantify the spin splitting near the valence-band edge, we evaluate the energy separation between the two uppermost valence bands of the $-P_z$ state where their $\langle\sigma_y\rangle$ projections have opposite signs (\textcolor{blue}{Fig.~S6}). Along the highlighted portions of $S$--$Y$--$\Gamma$, the splitting reaches $29.0$~meV without SOC and $29.6$~meV with SOC, remaining finite at $\Gamma$ in both cases. The similar magnitude and momentum dependence indicate that this band separation is only weakly affected by SOC.

\begin{figure}[!htp]
        \centering
        \includegraphics[width=1.0\linewidth]{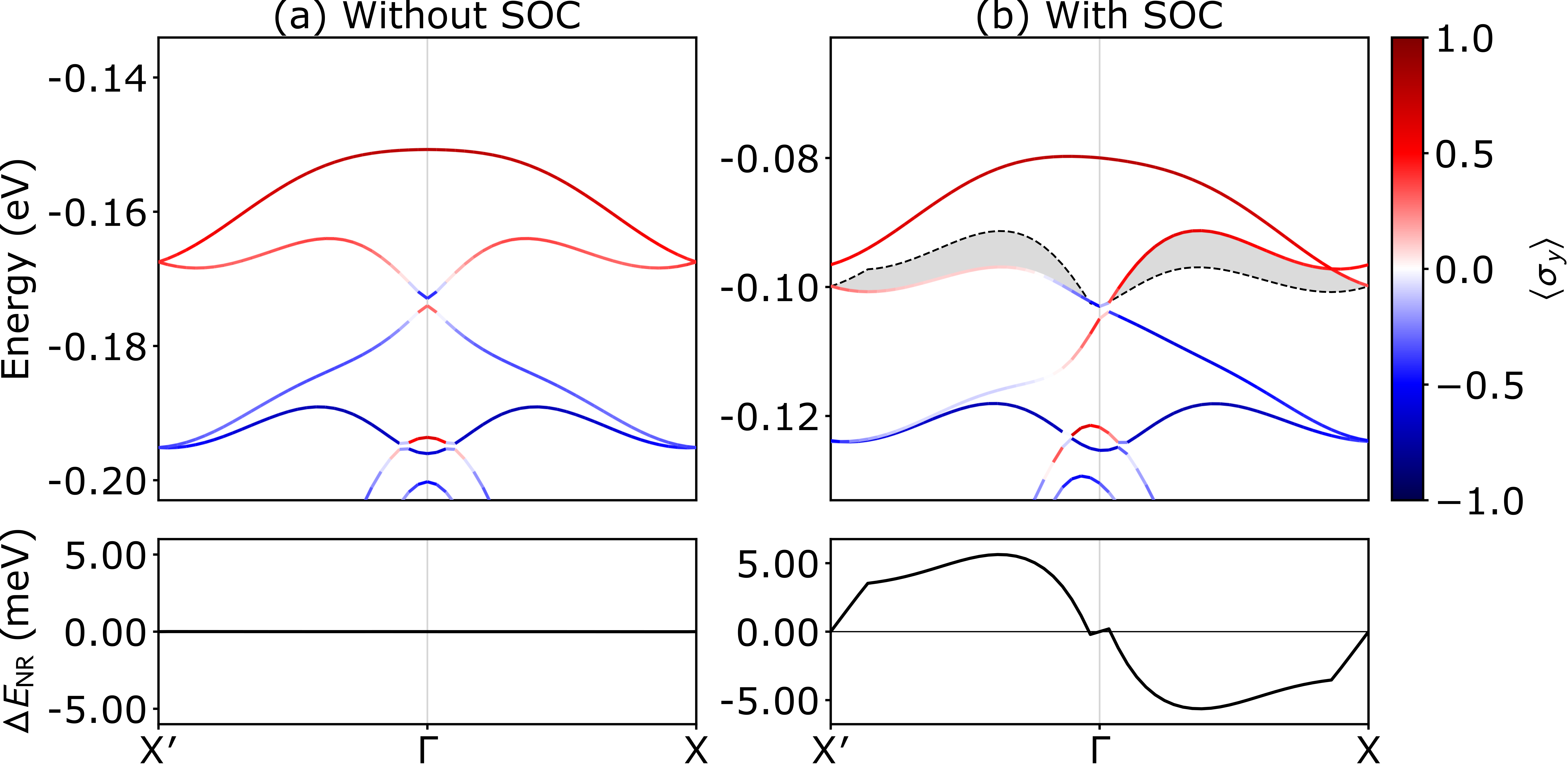}
        \caption{Reciprocal and nonreciprocal band dispersions of the $-P_z$ state calculated (a) without SOC and (b) with SOC, where $X'=(-\tfrac{1}{2},0,0)$. The color bar indicates the $y$-spin polarization $\langle\sigma_y\rangle$. The dashed curve in (b) is the second-highest valence band reflected about $\Gamma$; shading marks the deviation. Lower panels show $\Delta E_{\mathrm{NR}}(\mathbf{k})=E_n(-\mathbf{k})-E_n(\mathbf{k})$. The Fermi energy is set to zero.}
        \label{fig:reciprocity}
\end{figure}

Without SOC, the coplanar order preserves the antiunitary spin-only symmetry $[m_z^{\mathrm{s}}\thinspace\Vert\thinspace E]=C_{2z}^{\mathrm{s}}\mathcal{T}$, which leaves the in-plane moments unchanged and maps $\mathbf{k}$ to $-\mathbf{k}$, enforcing $E_n(\mathbf{k})=E_n(-\mathbf{k})$. SOC removes this independent spin symmetry. All operations of the magnetic point group $m^\prime m2^\prime$ preserve $k_x$, therefore allowing nonreciprocal dispersion along $X^\prime$--$\Gamma$--$X$. Consistent with this analysis, the calculated bands are reciprocal without SOC, whereas SOC produces an opposite-momentum energy difference of up to approximately $5$--$6$~meV (Fig.~\ref{fig:reciprocity}).

{\it Multiferroic--}
Since Ni$_3$B$_7$O$_{13}$Cl is an experimentally established ferroelectric that can be prepared in a single-domain state~\cite{haida1974,rivera1991linear}, we focus on the properties of the two polar states and do not address the switching pathway or barrier, which lies beyond the scope of this work. The crystallographic point group $mm2$ restricts the electric polarization to the $z$ axis. The mirror operations $m_{x}$ and $m_{y}$ reverse $P_x$ and $P_y$, respectively, while $C_{2z}$ reverses both components. Invariance under these operations therefore requires $P_x=P_y=0$, whereas $P_z$ remains symmetry-allowed. This symmetry-imposed form is consistent with our calculations, which yield $P_z=-0.85$ and $+0.85~\mu\mathrm{C/cm^2}$ for the $-P_z$ and $+P_z$ states, respectively (\textcolor{blue}{Fig.~S1}). 

The magnetic point group $m^\prime m2^\prime$ permits two independent linear magnetoelectric components, $\alpha_{yz}$ and $\alpha_{zy}$, defined here through the converse response $\Delta M_i=\alpha_{ij}E_j$. In particular, $\alpha_{yz}$ couples an electric field along the polar $z$ direction to a magnetization response along $y$. Both corresponding direct-response components have been measured in single-domain crystals below $T_N$~\cite{rivera1991linear}. Our calculations show that the $-P_z$ and $+P_z$ states exhibit reversed band-resolved spin polarizations $\langle\sigma_y\rangle$ while the net magnetization retains its direction, revealing the polar-state dependence of the electronic spin texture (\textcolor{blue}{Fig.~S7}). This comparison does not determine $\alpha_{yz}$; obtaining its magnitude would require a finite-electric-field response calculation.

\textit{Nonlinear photocurrents}--
As summarized in Table~\ref{tab:charge_photocurrent_components_niboracite_mpg}, the magnetic point group $m^\prime m2^\prime$ allows five independent real symmetric components of the linear SC tensor and two independent imaginary antisymmetric components of the circular IC tensor (\textcolor{blue}{Fig.~S8}). Both $\sigma_{zzz}$ and $\eta_{yyz}$ exhibit nearly equal magnitudes and opposite signs for the $-P_z$ and $+P_z$ states (\textcolor{blue}{Fig.~S9}). This ferroelectric-polarization-controlled reversal suggests potential applications in electrically programmable photodetection and nonvolatile memory with photovoltaic readout~\cite{gong2025reconfigurable,guo2013non}. The magnitude of $\sigma_{zzz}$ reaches $\approx1.1~\mu$A\,V$^{-2}$ near $3.7$~eV, close to a local maximum in the joint density of states (JDOS) (\textcolor{blue}{Fig.~S10}). A further correspondence between local maxima occurs near $4.9$~eV. This correspondence suggests that the availability of interband transitions contributes to the shift-current response in this energy range.

\begin{table}
\caption{Second-order charge-photocurrent tensor components allowed for Ni$_3$B$_7$O$_{13}$Cl. The shift and injection tensors are denoted by $\sigma_{ijk}$ and $\eta_{ijk}$, respectively, where $i$ labels the current direction and $j,k$ label the optical electric-field directions. For the antisymmetric tensors, minus-sign relations are shown explicitly. The symmetry transformations of all tensor components are provided in the SI (\textcolor{blue}{Table~S2 and S3}).}
\label{tab:charge_photocurrent_components_niboracite_mpg}

\centering
\renewcommand{\arraystretch}{1.20}
\setlength{\tabcolsep}{5pt}

\begin{tabular}{cccc}
\hline\hline
Channel
& Part
& \makecell{Jahn\\symbol}
& \makecell{Tensor\\components} \\
\hline

\makecell{Linear SC}
& \makecell{Real\\symmetric}
& $V[V^2]$
& \makecell[l]{
$\sigma_{zxx}$, $\sigma_{zyy}$, $\sigma_{zzz}$,\\
$\sigma_{xxz}$ = $\sigma_{xzx}$,\\
 $\sigma_{yyz}$ =  $\sigma_{yzy}$
} \\

\hline

\makecell{Circular IC}
& \makecell{Imaginary\\antisymmetric}
& $V\{V^2\}$
& \makecell[l]{
$\eta_{xxz}=-\eta_{xzx}$,\\
$\eta_{yyz}=-\eta_{yzy}$
} \\

\hline\hline
\end{tabular}
\end{table}

{\it Conclusions--}
We identify multiferroic Ni$_3$B$_7$O$_{13}$Cl as, to our knowledge, the first material hosting coexisting ALM and CFIM motifs in a single coplanar noncollinear magnetic state, a coexistence incompatible with collinear spin order. The ALM motif contributes a $d$-wave-like $\langle\sigma_x\rangle$ component, while the CFIM motif predominates in the electronic structure, producing $s$-wave-like spin polarization $\langle\sigma_y\rangle$. A minimal symmetry-allowed Hamiltonian accounts for the finite spin splitting at $\Gamma$, where the $d$-wave term vanishes while the $s$-wave contribution remains finite.  The reversal of band-resolved spin polarization between opposite ferroelectric states reveals magnetoelectric coupling, while sign reversals of the linear SC and circular IC responses demonstrate ferroelectric control of the photocurrent direction. These findings show that noncollinear multiferroics can exhibit coexisting ALM and CFIM motifs in real space and complementary spin textures in momentum space, highlighting them as viable candidates for spintronics and photovoltaic readout.

{\it Acknowledgments--} 
Harish K. Singh acknowledges the EuroHPC Joint Undertaking for granting access to the VEGA supercomputer at IZUM, Slovenia, through project EHPC-REG-2025R01-197-EuroHPC. The authors acknowledge the computing time provided on the Lichtenberg high-performance computer by the NHR Center NHR4CES at TU Darmstadt. V. Chaudhary was part of the NFDI consortium FAIRmat funded by Deutsche Forschungsgemeinschaft (DFG) - Project 460197019. Fu Li acknowledges support from the China Scholarship Council. H.~Zhang also acknowledges support from the Deutsche Forschungsgemeinschaft (DFG, German Research Foundation), Project-ID 463184206.

\textit{Data availability}---The data that support the findings of this work are available from the corresponding author upon reasonable request.
\bibliography{reference}

@article{blochl1994,
  title   = {Projector augmented-wave method},
  author  = {Bl{\"o}chl, P. E.},
  journal = {Phys. Rev. B},
  volume  = {50},
  issue   = {24},
  pages   = {17953--17979},
  numpages = {0},
  year    = {1994},
  month   = {Dec},
  publisher = {American Physical Society},
  doi     = {10.1103/PhysRevB.50.17953}
}

@article{kresse1999,
  title   = {From ultrasoft pseudopotentials to the projector augmented-wave method},
  author  = {Kresse, G. and Joubert, D.},
  journal = {Phys. Rev. B},
  volume  = {59},
  issue   = {3},
  pages   = {1758--1775},
  numpages = {0},
  year    = {1999},
  month   = {Jan},
  publisher = {American Physical Society},
  doi     = {10.1103/PhysRevB.59.1758},
  url     = {https://link.aps.org/doi/10.1103/PhysRevB.59.1758}
}

@article{kresse1993abinitio,
  title   = {Ab initio molecular dynamics for liquid metals},
  author  = {Kresse, G. and Hafner, J.},
  journal = {Phys. Rev. B},
  volume  = {47},
  issue   = {1},
  pages   = {558--561},
  numpages = {0},
  year    = {1993},
  month   = {Jan},
  publisher = {American Physical Society},
  doi     = {10.1103/PhysRevB.47.558},
  url     = {https://link.aps.org/doi/10.1103/PhysRevB.47.558}
}

@article{kresse1994abinitio,
  title   = {Ab initio molecular-dynamics simulation of the liquid-metal--amorphous-semiconductor transition in germanium},
  author  = {Kresse, G. and Hafner, J.},
  journal = {Phys. Rev. B},
  volume  = {49},
  issue   = {20},
  pages   = {14251--14269},
  numpages = {0},
  year    = {1994},
  month   = {May},
  publisher = {American Physical Society},
  doi     = {10.1103/PhysRevB.49.14251},
  url     = {https://link.aps.org/doi/10.1103/PhysRevB.49.14251}
}

@article{kresse1996efficient,
  title   = {Efficient iterative schemes for ab initio total-energy calculations using a plane-wave basis set},
  author  = {Kresse, Georg and Furthm{\"u}ller, J{\"u}rgen},
  journal = {Phys. Rev. B},
  volume  = {54},
  number  = {16},
  pages   = {11169},
  year    = {1996},
  doi={ https://doi.org/10.1103/PhysRevB.54.11169},
  publisher = {APS}
}

@article{kresse1996efficiency,
  title   = {Efficiency of ab-initio total energy calculations for metals and semiconductors using a plane-wave basis set},
  author  = {Kresse, Georg and Furthm{\"u}ller, J{\"u}rgen},
  journal = {Comput. Mater. Sci.},
  volume  = {6},
  number  = {1},
  pages   = {15--50},
  year    = {1996},
  doi={https://doi.org/10.1016/0927-0256(96)00008-0},
  publisher = {Elsevier}
}

@article{perdew1996generalized,
  title   = {Generalized gradient approximation made simple},
  author  = {Perdew, John P and Burke, Kieron and Ernzerhof, Matthias},
  journal = {Phys. Rev. Lett.},
  volume  = {77},
  number  = {18},
  pages   = {3865},
  year    = {1996},
  doi={https://doi.org/10.1103/PhysRevLett.77.3865},
  publisher = {APS}
}

@article{ye1997neutron,
  title={A Neutron Diffraction Study of the Magnetic Structure and Phase Transition in {Ni$_3$B$_7$O$_{13}$Cl} Boracite},
  author={Ye, Zuo-Guang and Schobinger-Papamantellos, Penelope and Mao, Shao Yu and Ritter, C and Suard, E and Sato, M and Schmid, Hans},
  journal={Ferroelectrics},
  volume={204},
  number={1},
  pages={83--96},
  year={1997},
  doi={https://doi.org/10.1080/00150199708222190},
  publisher={New York: Gordon and Breach, c1970-}
}

@article{hayami2019momentum,
  title={Momentum-dependent spin splitting by collinear antiferromagnetic ordering},
  author={Hayami, Satoru and Yanagi, Yuki and Kusunose, Hiroaki},
  journal={J. Phys. Soc. Jpn.},
  volume={88},
  number={12},
  pages={123702},
  year={2019},
  doi={https://doi.org/10.7566/JPSJ.88.123702},
  publisher={The Physical Society of Japan}
}

@article{vsmejkal2020crystal,
  title={Crystal time-reversal symmetry breaking and spontaneous {H}all effect in collinear antiferromagnets},
  author={{\v{S}}mejkal, Libor and Gonz{\'a}lez-Hern{\'a}ndez, Rafael and Jungwirth, Tom{\'a}{\v{s}} and Sinova, Jairo},
  journal={Sci. Adv.},
  volume={6},
  number={23},
  pages={eaaz8809},
  year={2020},
  doi={10.1126/sciadv.aaz8809},
  publisher={American Association for the Advancement of Science}
}

@article{yuan2021prediction,
  title   = {Prediction of low-Z collinear and noncollinear antiferromagnetic compounds having momentum-dependent spin splitting even without spin-orbit coupling},
  author  = {Yuan, Lin-Ding and Wang, Zhi and Luo, Jun-Wei and Zunger, Alex},
  journal = {Phys. Rev. Mater.},
  volume  = {5},
  number  = {1},
  pages   = {014409},
  year    = {2021},
  doi={https://doi.org/10.1103/PhysRevMaterials.5.014409},
  publisher = {APS}
}

@article{mazin2021prediction,
  title={Prediction of unconventional magnetism in doped {FeSb$_2$}},
  author={Mazin, Igor I and Koepernik, Klaus and Johannes, Michelle D and Gonz{\'a}lez-Hern{\'a}ndez, Rafael and {\v{S}}mejkal, Libor},
  journal={Proc. Natl. Acad. Sci. U.S.A.},
  volume={118},
  number={42},
  pages={e2108924118},
  year={2021},
  doi={https://doi.org/10.1073/pnas.2108924118},
  publisher={National Academy of Sciences}
}

@article{guo2023spin,
  title={Spin-split collinear antiferromagnets: A large-scale ab-initio study},
  author={Guo, Yaqian and Liu, Hui and Janson, Oleg and Fulga, Ion Cosma and van den Brink, Jeroen and Facio, Jorge I},
  journal={Mater. Today Phys.},
  volume={32},
  pages={100991},
  year={2023},
  doi={https://doi.org/10.1016/j.mtphys.2023.100991},
  publisher={Elsevier}
}

@article{vsmejkal2022beyond,
  title   = {Beyond conventional ferromagnetism and antiferromagnetism: A phase with nonrelativistic spin and crystal rotation symmetry},
  author  = {{\v{S}}mejkal, Libor and Sinova, Jairo and Jungwirth, Tomas},
  journal = {Phys. Rev. X},
  volume  = {12},
  number  = {3},
  pages   = {031042},
  year    = {2022},
  doi={https://doi.org/10.1103/PhysRevX.12.031042},
  publisher = {APS}
}

@article{krempasky2024altermagnetic,
  title={Altermagnetic lifting of {K}ramers spin degeneracy},
  author={Krempask{\`y}, J and {\v{S}}mejkal, L and D’souza, SW and Hajlaoui, M and Springholz, G and Uhl{\'\i}{\v{r}}ov{\'a}, K and Alarab, F and Constantinou, PC and Strocov, V and Usanov, D and others},
  journal={Nature},
  volume={626},
  number={7999},
  pages={517--522},
  year={2024},
  doi={https://doi.org/10.1038/s41586-023-06907-7},
  publisher={Nature Publishing Group UK London}
}

@article{reimers2024direct,
  title={Direct observation of altermagnetic band splitting in {CrSb} thin films},
  author={Reimers, Sonka and Odenbreit, Lukas and {\v{S}}mejkal, Libor and Strocov, Vladimir N and Constantinou, Procopios and Hellenes, Anna B and Jaeschke Ubiergo, Rodrigo and Campos, Warlley H and Bharadwaj, Venkata K and Chakraborty, Atasi and others},
  journal={Nat. Commun.},
  volume={15},
  number={1},
  pages={2116},
  year={2024},
  doi={https://doi.org/10.1038/s41467-024-46476-5},
  publisher={Nature Publishing Group UK London}
}

@article{gonzalez2021efficient,
  title={Efficient electrical spin splitter based on nonrelativistic collinear antiferromagnetism},
  author={Gonz{\'a}lez-Hern{\'a}ndez, Rafael and {\v{S}}mejkal, Libor and V{\`y}born{\`y}, Karel and Yahagi, Yuta and Sinova, Jairo and Jungwirth, Tom{\'a}{\v{s}} and {\v{Z}}elezn{\`y}, Jakub},
  journal={Phys. Rev. Lett.},
  volume={126},
  number={12},
  pages={127701},
  year={2021},
  doi={https://doi.org/10.1103/PhysRevLett.126.127701},
  publisher={APS}
}

@article{yuan2024nonrelativistic,
  title   = {Nonrelativistic spin splitting at the {B}rillouin zone center in compensated magnets},
  author  = {Yuan, Lin-Ding and Georgescu, Alexandru B and Rondinelli, James M},
  journal = {Phys. Rev. Lett.},
  volume  = {133},
  number  = {21},
  pages   = {216701},
  year    = {2024},
  doi={https://doi.org/10.1103/PhysRevLett.133.216701},
  publisher = {APS}
}

@article{guo2025luttinger,
  title={{L}uttinger-compensated bipolarized magnetic semiconductor},
  author={Guo, Peng-Jie and Yang, Huan-Cheng and Hou, Xiao-Yao and Gao, Ze-Feng and Ji, Wei and Lu, Zhong-Yi},
  journal={Phys. Rev. B},
  volume={112},
  number={18},
  pages={L180405},
  year={2025},
  doi={https://doi.org/10.1103/9syc-71w8},
  publisher={APS}
}

@article{guo2025achieving,
    author = {Guo, San-Dong and He, Junjie and Ang, Yee Sin},
    title = {Achieving fully compensated ferrimagnetism through two-dimensional {CrI$_3$/CrGeTe$_3$} heterojunctions},
    journal = {Appl. Phys. Lett.},
    volume = {127},
    number = {23},
    pages = {232401},
    year = {2025},
    month = {12},
    issn = {0003-6951},
    doi = {10.1063/5.0303560},
    url = {https://doi.org/10.1063/5.0303560}
}

@article{liu2025two,
  title={Two-dimensional fully compensated ferrimagnetism},
  author={Liu, Yichen and Guo, San-Dong and Li, Yongpan and Liu, Cheng-Cheng},
  journal={Phys. Rev. Lett.},
  volume={134},
  number={11},
  pages={116703},
  year={2025},
  doi={https://doi.org/10.1103/PhysRevLett.134.116703},
  publisher={APS}
}

@article{dong2025nonrelativistic,
  title={Nonrelativistic spin-splitting multiferroic antiferromagnets and compensated ferrimagnet with zero net magnetization},
  author={Dong, Jianting and Wu, Kun and Zhu, Meng and Zheng, Fanxing and Li, Xinlu and Zhang, Jia},
  journal={Phys. Rev. B},
  volume={112},
  number={2},
  pages={024425},
  year={2025},
  doi={https://doi.org/10.1103/74sw-w1fv},
  publisher={APS}
}

@article{zhang2025fully,
  title={Fully compensated ferrimagnetism in a {Cr$_2$TiC$_2$} monolayer {MX}ene},
  author={Zhang, Zhiwei and Wei, Yajie and Zhang, Dingwen and Ye, Haoshen and Zhang, GP and Bai, Dongmei and Wang, Jianli},
  journal={Nanoscale},
  volume={17},
  number={39},
  pages={22898--22904},
  year={2025},
  doi={https://doi.org/10.1039/D5NR02850G},
  publisher={Royal Society of Chemistry}
}

@article{liu2025robust,
  title={Robust altermagnetism and compensated ferrimagnetism in {MnPX$_3$}-based {(X= S or Se)} heterostructures},
  author={Liu, Yunsong and Liu, Yanlong and Wang, Xuefei and Xia, Nan and Xu, Guifang and Wang, Yi and Wang, Haifeng and Gao, Weiwei and Zhao, Jijun},
  journal={Phys. Rev. Mater.},
  volume={9},
  number={10},
  pages={104406},
  year={2025},
  doi={https://doi.org/10.1103/6wmg-hxdq},
  publisher={APS}
}

@article{guo2026electric,
    author = {Guo, Deping and Zhang, Weihan and Zong, Canbo and Wang, Cong and Ji, Wei},
    title = {Electric field-tunable {L}uttinger compensated antiferromagnetism in {CrCl$_2$} double chains},
    journal = {Appl. Phys. Lett.},
    volume = {128},
    number = {11},
    pages = {111602},
    year = {2026},
    month = {03},
    doi = {10.1063/5.0305548}
}

@article{guo2026route,
  title={A route to fully-compensated ferrimagnetic metal: electric-field annihilation of the bilayer bandgap},
  author={Guo, San-Dong and Bian, Rongyuan and Fan, Feng-Ren and Stroppa, Alessandro},
  journal={J. Phys.: Condens. Matter},
  volume={38},
  number={22},
  pages={225801},
  year={2026},
  doi={10.1088/1361-648X/ae72dd}
}

@article{guo2026external,
  title={External field induced transition from altermagnetic metal to fully compensated ferrimagnetic metal in monolayer {Cr$_2$O}},
  author={Guo, San-Dong and Luo, Qiqi and Zhang, Shi-Hao and Jiang, Peng},
  journal={Phys. Rev. B},
  volume={113},
  number={6},
  pages={064408},
  year={2026},
  doi={https://doi.org/10.1103/qnwp-8sx5},
  publisher={APS}
}

@article{aribot2026persistent,
  title={Persistent compensated ferrimagnetism in the molecular framework {Cr(pyrazine)$_3$}},
  author={Aribot, Fr{\'e}d{\'e}ric and Dunstan, Maja A and Yutronkie, Nathan J and Kubus, Mariusz and Leyva-Parra, Luis and Viborg, Anton and Mossin, Susanne and Wilhelm, Fabrice and Nassif, Vivian and Pinkowicz, Dawid and others},
  journal={Nat. Chem.},
  pages={1366--1371},
  number=[8],
  volume={18},
  doi={https://doi.org/10.1038/s41557-026-02131-8},
  year={2026}
}

@article{zhou2021efficient,
  title={Efficient spintronics with fully compensated ferrimagnets},
  author={Zhou, Heng-An and Xu, Teng and Bai, Hao and Jiang, Wanjun},
  journal={J. Phys. Soc. Jpn.},
  volume={90},
  number={8},
  pages={081006},
  year={2021},
  doi={https://doi.org/10.7566/JPSJ.90.081006},
  publisher={The Physical Society of Japan}
}

@article{finley2020spintronics,
    author = {Finley, Joseph and Liu, Luqiao},
    title = {Spintronics with compensated ferrimagnets},
    journal = {Appl. Phys. Lett.},
    volume = {116},
    number = {11},
    pages = {110501},
    year = {2020},
    month = {03},
    issn = {0003-6951},
    doi = {10.1063/1.5144076},
    url = {https://doi.org/10.1063/1.5144076}
}

@article{semboshi2022new,
  title={A new type of half-metallic fully compensated ferrimagnet},
  author={Semboshi, Satoshi and Umetsu, Rie Y and Kawahito, Yousuke and Akai, Hisazumi},
  journal={Sci. Rep.},
  volume={12},
  number={1},
  pages={10687},
  year={2022},
  doi={https://doi.org/10.1038/s41598-022-14561-8},
  publisher={Nature Publishing Group UK London}
}

@article{lee2025spin,
  title={Spin transport phenomena in junctions composed of a compensated ferrimagnet and a normal metal},
  author={Lee, Xin Theng and Misawa, Takahiro and Matsuo, Mamoru and Kato, Takeo},
  journal={arXiv preprint arXiv:2507.05618},
  url={https://arxiv.org/abs/2507.05618},
  year={2025}
}

@article{zhang2025prototypes,
  title={Prototypes of Nonrelativistic Spin Splitting and Polarization in Symmetry Broken Antiferromagnets},
  author={Zhang, Xiuwen and Xiong, Jia-Xin and Yuan, Lin-Ding and Zunger, Alex},
  journal={Phys. Rev. X},
  volume={15},
  number={3},
  pages={031076},
  year={2025},
  doi={https://doi.org/10.1103/mrzv-wmcf},
  publisher={APS}
}

@article{kawamura2024compensated,
  title={Compensated ferrimagnets with colossal spin splitting in organic compounds},
  author={Kawamura, Taiki and Yoshimi, Kazuyoshi and Hashimoto, Kenichiro and Kobayashi, Akito and Misawa, Takahiro},
  journal={Phys. Rev. Lett.},
  volume={132},
  number={15},
  pages={156502},
  year={2024},
  doi={https://doi.org/10.1103/PhysRevLett.132.156502},
  publisher={APS}
}

@article{park2022anomalous,
  title={Anomalous {H}all effect in a compensated ferrimagnet: Symmetry analysis for {Mn$_3$Al}},
  author={Park, Minkyu and Han, Guihyun and Rhim, SH},
  journal={Phys. Rev. Res.},
  volume={4},
  number={1},
  pages={013215},
  year={2022},
  doi={https://doi.org/10.1103/PhysRevResearch.4.013215},
  publisher={APS}
}

@article{han2026strain,
  title={Strain tunable anomalous {H}all and Nernst conductivities in compensated ferrimagnetic {Mn$_3$Al}},
  author={Han, Guihyun and Park, Minkyu and Rhim, SH},
  journal={Phys. Rev. B},
  volume={113},
  number={9},
  pages={094425},
  year={2026},
  doi={https://doi.org/10.1103/wgzp-3b9n},
  publisher={APS}
}

@article{hu2026non,
  title={Non-collinear Altermagnetic Phases in the Mott Insulator {NiS$_2$}},
  author={Hu, Mengli and Iraola, Mikel I and McClarty, Paul and Brink, Jeroen van den and Vergniory, Maia G},
  journal={arXiv preprint arXiv:2603.01329},
  url={http://10.48550/arXiv.2603.01329},
  year={2026}
}

@article{cheong2024altermagnetism,
  title={Altermagnetism with non-collinear spins},
  author={Cheong, Sang-Wook and Huang, Fei-Ting},
  journal={npj Quantum Mater.},
  volume={9},
  number={1},
  pages={13},
  year={2024},
  doi={https://doi.org/10.1038/s41535-024-00626-6},
  publisher={Nature Publishing Group UK London}
}

@article{hu2025spin,
  title={Spin {H}all and Edelstein effects in chiral non-collinear altermagnets},
  author={Hu, Mengli and Janson, Oleg and Felser, Claudia and McClarty, Paul and Van den Brink, Jeroen and G. Vergniory, Maia},
  journal={Nat. Commun.},
  volume={16},
  number={1},
  pages={8529},
  year={2025},
  doi={https://doi.org/10.1038/s41467-025-64271-8},
  publisher={Nature Publishing Group UK London}
}

@article{eerenstein2006multiferroic,
  title={Multiferroic and magnetoelectric materials},
  author={Eerenstein, Wilma and Mathur, ND and Scott, James F},
  journal={Nature},
  volume={442},
  number={7104},
  pages={759--765},
  year={2006},
  doi={https://doi.org/10.1038/nature05023},
  publisher={Nature Publishing Group UK London}
}

@article{zhao2026recent,
  title={Recent Advances of Bulk Photovoltaic Effect in Exotic Quantum Materials: Progress and Challenges},
  author={Zhao, Yiwei and Xue, Qianqian and Mu, Xingchi and Cui, Hanli and Zhou, Jian},
  journal={Adv. Mater.},
  volume = {38},
  number = {29},
  year={2026},
  pages = {e17783},
  doi={https://doi.org/10.1002/adma.202517783},
  publisher={Wiley Online Library}
}

@article{kim2017shift,
  title={Shift charge and spin photocurrents in {D}irac surface states of topological insulator},
  author={Kim, Kun Woo and Morimoto, Takahiro and Nagaosa, Naoto},
  journal={Phys. Rev. B},
  volume={95},
  number={3},
  pages={035134},
  year={2017},
  doi={ https://doi.org/10.1103/PhysRevB.95.035134},
  publisher={APS}
}

@article{panday2019injection,
  title={Injection current in ferroelectric group-{IV} monochalcogenide monolayers},
  author={Panday, Suman Raj and Barraza-Lopez, Salvador and Rangel, Tonatiuh and Fregoso, Benjamin M},
  journal={Phys. Rev. B},
  volume={100},
  number={19},
  pages={195305},
  year={2019},
  doi={https://doi.org/10.1103/PhysRevB.100.195305},
  publisher={APS}
}

@article{jiang2025nonlinear,
  title={Nonlinear photocurrent as a hallmark of altermagnet},
  author={Jiang, Xiao and Jeong, Uiseok and Sato, Shunsuke and Shin, Dongbin and Yabana, Kazuhiro and Yan, Binghai and Park, Noejung},
  journal={ACS Nano},
  volume={19},
  number={26},
  pages={23620--23628},
  year={2025},
  publisher={ACS Publications},
  doi={https://doi.org/10.1021/acsnano.5c01421?urlappend=%3Fref%3DPDF&jav=VoR&rel=cite-as}
}

@article{ezawa2025bulk,
  title     = {Bulk photovoltaic effects in altermagnets},
  author    = {Ezawa, Motohiko},
  journal   = {Phys. Rev. B},
  volume    = {111},
  issue     = {20},
  pages     = {L201405},
  numpages  = {7},
  year      = {2025},
  month     = {May},
  publisher = {American Physical Society},
  doi       = {10.1103/PhysRevB.111.L201405},
  url       = {https://doi.org/10.1103/PhysRevB.111.L201405}
}

@article{dong2025crystal,
  title     = {Crystal symmetry selected pure spin photocurrent in altermagnetic insulators},
  author    = {Dong, Ruizhi and Cao, Ranquan and Tan, Dian and Fei, Ruixiang},
  journal   = {Phys. Rev. B},
  volume    = {111},
  issue     = {19},
  pages     = {195210},
  numpages  = {7},
  year      = {2025},
  month     = {May},
  publisher = {American Physical Society},
  doi       = {10.1103/PhysRevB.111.195210},
  url       = {https://doi.org/10.1103/PhysRevB.111.195210}
}

@article{haida1974,
author = {Haida ,Munetaka and Kohn ,Kay and Schmid ,Hans},
title = {The Magnetic Properties of Ferroelectric and Ferromagnetic {Ni-Cl} Boracite at Low Temperature},
journal = {J. Phys. Soc. Jpn.},
volume = {37},
number = {5},
pages = {1463-1463},
year = {1974},
doi = {10.1143/JPSJ.37.1463},
url = {https://doi.org/10.1143/JPSJ.37.1463}
}

@article{rivera1991linear,
  title={Linear and quadratic magnetoelectric ({ME}) effect in {Ni-Cl} boracite},
  author={Rivera, J-P and Schmid, Hans},
  journal={J. Appl. Phys.},
  volume={70},
  number={10},
  pages={6410--6412},
  year={1991},
  doi={https://doi.org/10.1063/1.349911},
  publisher={American Institute of Physics}
}

@article{gong2025reconfigurable,
  title={Reconfigurable and nonvolatile ferroelectric bulk photovoltaics based on {3R-WS$_2$} for machine vision},
  author={Gong, Yue and Duan, Ruihuan and Hu, Yi and Wu, Yao and Zhu, Song and Wang, Xingli and Wang, Qijie and Lau, Shu Ping and Liu, Zheng and Tay, Beng Kang},
  journal={Nat. Commun.},
  volume={16},
  number={1},
  pages={230},
  year={2025},
  doi={https://doi.org/10.1038/s41467-024-55562-7},
  publisher={Nature Publishing Group UK London}
}

@article{guo2013non,
  title={Non-volatile memory based on the ferroelectric photovoltaic effect},
  author={Guo, Rui and You, Lu and Zhou, Yang and Shiuh Lim, Zhi and Zou, Xi and Chen, Lang and Ramesh, R and Wang, Junling},
  journal={Nat. Commun.},
  volume={4},
  number={1},
  pages={1990},
  year={2013},
  doi={https://doi.org/10.1038/ncomms2990},
  publisher={Nature Publishing Group UK London}
}

@article{li2026medium,
  title={Medium-Throughput Evaluation of Quantum Geometry-Driven Topological Transports in Altermagnets},
  author={Li, Fu and Zhao, Bo and Chaudhary, Vikrant and Wang, Shengqiao and Shen, Chen and Wang, Hao and Zhang, Hongbin},
  journal={arXiv preprint arXiv:2604.17071},
  url={http://10.48550/arXiv.2604.17071},
  year={2026}
}

@article{vsmejkal2022emerging,
  title={Emerging research landscape of altermagnetism},
  author={{\v{S}}mejkal, Libor and Sinova, Jairo and Jungwirth, Tomas},
  journal={Phys. Rev. X},
  volume={12},
  number={4},
  pages={040501},
  year={2022},
  doi={https://doi.org/10.1103/PhysRevX.12.040501},
  publisher={APS}
}

@article{bai2024altermagnetism,
  title={Altermagnetism: Exploring new frontiers in magnetism and spintronics},
  author={Bai, Ling and Feng, Wanxiang and Liu, Siyuan and {\v{S}}mejkal, Libor and Mokrousov, Yuriy and Yao, Yugui},
  journal={Adv. Funct. Mater.},
  volume={34},
  number={49},
  pages={2409327},
  year={2024},
  doi={https://doi.org/10.1002/adfm.202409327},
  publisher={Wiley Online Library}
}

@article{gu2025ferroelectric,
  title={Ferroelectric switchable altermagnetism},
  author={Gu, Mingqiang and Liu, Yuntian and Zhu, Haiyuan and Yananose, Kunihiro and Chen, Xiaobing and Hu, Yongkang and Stroppa, Alessandro and Liu, Qihang},
  journal={Phys. Rev. Lett.},
  volume={134},
  number={10},
  pages={106802},
  year={2025},
  doi={https://doi.org/10.1103/PhysRevLett.134.106802},
  publisher={APS}
}

@article{schnelle2015magnetic,
  title={Magnetic and structural phase transitions of multiferroic boracites {M$_3$B$_7$O$_1$$_3$X} {(M= 3 d transition metal {Cr--Zn} or {Mg; X}= halogen {Cl, Br, I})}},
  author={Schnelle, Walter and Schmid, Hans},
  journal={Phys. Rev. B},
  volume={91},
  number={18},
  pages={184411},
  year={2015},
  doi={https://doi.org/10.1103/PhysRevB.91.184411},
  publisher={APS}
}

@article{stanciu2006ultrafast,
  title     = {Ultrafast Spin Dynamics across Compensation Points in
               Ferrimagnetic {GdFeCo}: The Role of Angular Momentum
               Compensation},
  author    = {Stanciu, C. D. and Kimel, A. V. and Hansteen, F. and
               Tsukamoto, A. and Itoh, A. and Kirilyuk, A. and Rasing, Th.},
  journal   = {Phys. Rev. B},
  volume    = {73},
  number    = {22},
  pages     = {220402(R)},
  year      = {2006},
  doi       = {10.1103/PhysRevB.73.220402},
  publisher = {American Physical Society}
}

@article{kim2017fast,
  title   = {Fast Domain Wall Motion in the Vicinity of the Angular
             Momentum Compensation Temperature of Ferrimagnets},
  author  = {Kim, Kab-Jin and Kim, Se Kwon and Hirata, Yuushou and
             Oh, Se-Hyeok and Tono, Takayuki and Kim, Duck-Ho and
             Okuno, Takaya and Ham, Woo Seung and Kim, Sanghoon and
             Go, Gyoungchoon and Tserkovnyak, Yaroslav and
             Tsukamoto, Arata and Moriyama, Takahiro and
             Lee, Kyung-Jin and Ono, Teruo},
  journal = {Nat. Mater.},
  volume  = {16},
  number  = {12},
  pages   = {1187--1192},
  year    = {2017},
  doi     = {10.1038/nmat4990}
}

@article{caretta2018fast,
  title   = {Fast Current-Driven Domain Walls and Small Skyrmions in a
             Compensated Ferrimagnet},
  author  = {Caretta, Lucas and Mann, Maxwell and B{\"u}ttner, Felix and
             Ueda, Kohei and Pfau, Bastian and G{\"u}nther, Christian M. and
             Hessing, Piet and Churikova, Alexandra and Klose, Christopher and
             Schneider, Michael and Engel, Dieter and Marcus, Colin and
             Bono, David and Bagschik, Kai and Eisebitt, Stefan and
             Beach, Geoffrey S. D.},
  journal = {Nat. Nanotechnol.},
  volume  = {13},
  number  = {12},
  pages   = {1154--1160},
  year    = {2018},
  doi     = {10.1038/s41565-018-0255-3}
}

@article{mavragani2026compensated,
author = {Niki Mavragani  and Alexandros A. Kitos  and Frédéric Aribot  and Nathan J. Yutronkie  and Thomas Veile  and Adam F. Sapnik  and Anton Viborg  and Luis Leyva-Parra  and Lukas Woodcock  and Thomas Krøier Rønne-Nielsen  and Maja A. Dunstan  and Mariusz Kubus  and Poul Norby  and Fabrice Wilhelm  and Jens Wenzel Andreasen  and Cathrine Frandsen  and Jesper Bendix  and Susanne Mossin  and Stergios Piligkos  and Sebastian E. Reyes-Lillo  and Andrei Rogalev  and Kasper S. Pedersen },
title = {Compensated ferrimagnetism above 600 {K} in a metal-organic framework},
journal = {ChemRxiv},
volume = {2026},
number = {0714},
pages = {},
year = {2026},
doi = {10.26434/chemrxiv.15006083/v1}
}

@article{li2026MagneticPhotocurrents,
  author        = {Li, Fu and Ray, Rajyavardhan and Liu, Xiaoxiong
                   and Prasad, Babu Baijnath and Shen, Chen
                   and Guo, Yaqian and van den Brink, Jeroen
                   and Zhang, Hongbin and Singh, Harish K.},
  title         = {Magnetic-Order-Driven Nonlinear Photocurrents
                   in {$\mathcal{PT}$}-Broken Antiferromagnetic
                   {MnS$_2$} and Altermagnetic {MnSe$_2$}},
  journal       = {arXiv preprint arXiv:2610.03450},
  year          = {2026},
  url           = {https://arxiv.org/abs/2610.03450v1}
}

@article{wang2026strain,
  title={Strain-Tunable Shift Current and Magneto-Optical Kerr Effect in Multiferroic Altermagnet {Fe$_2$Mo$_3$O$_8$}},
  author={Wang, Shengqiao and Zhao, Bo and Singh, Harish K and Xie, Jiahao and Li, Fu and Zhang, Hongbin and Su, Yang and Shen, Chen and Zhang, Lijun},
  journal={arXiv preprint arXiv:2607.12799},
  url={http://arxiv.org/abs/2607.12799},
  year={2026}
}
 
\end{document}